# A thermodynamic basis for prebiotic amino acid synthesis and the nature of the first genetic code

Paul G. Higgs and Ralph E. Pudritz

Origins Institute and Department of Physics and Astronomy,
McMaster University, Hamilton, Ontario.

Contact - higgsp@mcmaster.ca;   pudritz@mcmaster.ca

Running title - Prebiotic amino acid synthesis

**Abstract**

Of the twenty amino acids used in proteins, ten were formed in Miller's atmospheric discharge experiments. The two other major proposed sources of prebiotic amino acid synthesis include formation in hydrothermal vents and delivery to Earth via meteorites. We combine observational and experimental data of amino acid frequencies formed by these diverse mechanisms and show that, regardless of the source, these ten early amino acids can be ranked in order of decreasing abundance in prebiotic contexts. This order can be predicted by thermodynamics. The relative abundances of the early amino acids were most likely reflected in the composition of the first proteins at the time the genetic code originated. The remaining amino acids were incorporated into proteins after pathways for their biochemical synthesis evolved. This is consistent with theories of the evolution of the genetic code by stepwise addition of new amino acids. These are hints that key aspects of early biochemistry may be universal.

**Introduction - Possible Mechanisms of Prebiotic Amino Acid Synthesis**

Among the organic molecules that are constituents of cells, amino acids stand out by virtue of their role as building blocks of proteins and their ease of chemical formation. There has been clear evidence for prebiotic formation of amino acids since the experiments of Miller (Miller, 1953; Miller and Orgel, 1974) involving electrical discharge in a mixture of atmospheric gases, which produced 10 of the amino acids used in modern proteins plus many other organic molecules. Formation of prebiotic amino acids in interstellar gas, as well as in hydrothermal vents, has since been proposed and tested. We will show that these diverse environments are remarkably consistent about which amino acids formed.

The distinction between early and late amino acids is important for theories of the origin of the genetic code (*i.e.* the mapping between codons and amino acids – a codon is the combination of 3 bases in a nucleic acid sequence needed to specify an amino acid). Wong (1975; 2005) argues that early versions of the genetic code encoded only a few amino acids and that later amino acids were added as biosynthetic pathways evolved. With a view to determining the order of addition of amino acids to the genetic code, Trifonov (2004) collected 60 criteria by which the amino acids can be ranked. The consensus order that emerges from these criteria appears to be a useful one; however, we have reservations because diverse criteria were included that are not all consistent with one another. Our analysis below uses a similar ranking procedure but incorporates only criteria based on measurable amino acid concentrations.

First, let us consider astrophysical sites for the formation of organic molecules. Interstellar molecular cloud cores and the protostellar disks that form from their collapse are a universal aspect of star and planet formation. Observations of disks around other stars reveal that active chemistry occurs near the surface regions of protoplanetary disks (Bergin *et al.* 2007). Less volatile gases, such as CO, $NH_3$ and HCN, freeze out on dust grains and organic molecules can be synthesized on the surface of the grains after exposure to ultra-violet radiation. Alternatively, it has been proposed that amino acids can form during the gravitational collapse of a core, and predictions have been made regarding molecular concentrations and locations within the disk (Chakrabarti and Chakrabarti, 2000). Molecules in protoplanetary disks end up in comets and meteorites as the dust grains coalesce and are then delivered to Earth when impact events occur. Meteorites forming in the icy outer regions of the solar system can later impact the Earth if their orbits are perturbed by large outer planets. This provides a way of bringing organic molecules and water to Earth that formed on grains and in meteorites at larger orbital radii - beyond the snow lines for these ices (for water, typically at or beyond 2.5 Astronomical Units). Delivery of molecules from outside Earth is only relevant for the origin of life if the rate is comparable to or exceeds the rate of formation of the same molecules by Earth-based chemistry. Several authors have tried to estimate the relative rates of these internal and external processes, although it is extremely difficult to do this with certainty (Whittet, 1997; Pierazzo and Chyba, 1999; Bernstein, 2006). Observation of the chemical composition of meteorites provides us with a time capsule to the kind of organic chemistry that was occurring in the early stages of the solar system when the meteorite parent bodies were formed. In contrast, chemistry on Earth has been dramatically changed by subsequent geological and biological activity. Thus, meteoritic compositional data and extrapolated impact rate are relevant for understanding pre-biotic conditions and suggest that meteorites contributed significantly to the organics present on Earth over 4 GA ago - at delivery rates of the order $10^8$ kg per yr (Pierazzo and Chyba, 1999).

Table 1 gives the relative concentrations of amino acids measured in meteorites. Column M1 is the Murchison meteorite ($H_2O$ extract from Table 2 of Engel and Nagy, 1982), M2 is the Murray meteorite (Cronin and Moore, 1971), and M3 is the Yamato meteorite (interior, hydrolysed sample from Shimoyama *et al.* 1979). The possibility that dust grains with icy mantles are a site for amino acid synthesis has been investigated experimentally (Bernstein *et al.* 2002; Munoz Caro *et al.* 2002). Several amino acids may result from the exposure of icy mantles of HCN, $NH_3$, and $H_2CO$ to UV radiation. Column I1 shows measurements from Munoz Caro *et al.* (2002).

We next turn to Miller's experiments, which showed that synthesis was possible in the atmosphere of the early Earth - presumed to be reducing. Column A1 shows the results with an atmosphere of $CH_4$, $NH_3$, $H_2O$ and $H_2$ (Miller and Orgel, 1972), and column A2 shows the results with an atmosphere of $CH_4$, $N_2$, $H_2O$ and traces of $NH_3$ (Miller and Orgel, 1972). There have since been many claims that the early atmosphere was not strongly reducing (Kasting, 1993; Pierazzo and Chyba, 1999), but the importance of atmospheric synthesis is still maintained by some. Schlesinger and Miller (1983) have compared the quantities and the diversity of amino acids formed from various atmospheric mixtures. Although yields are lower in non-reducing atmospheres, synthesis is still possible in mixtures that are not strongly reducing, such as the experiment using proton irradiation of an atmosphere of CO, $N_2$ and $H_2O$ (Miyakawa *et al.* 2002), the results of which are shown in column A3. Another very recent result (Cleaves *et al.* 2008) on spark discharge in neutral gas mixtures dominated by $CO_2$ and $N_2$ shows that yields of amino acids in these conditions may be higher than previously estimated. Interestingly, two recent calculations regarding the atmosphere on early Earth (Tian *et al.* 2005; Schaefer and Fegley, 2007) claim that it may have been reducing after all and that the rates of organic synthesis may in fact be high.

In the experiments above, amino acids are formed by Strecker synthesis (Miller, 1957), *i.e.* energy sources such as spark discharge or UV produce HCN and aldehydes, which dissolve in water and subsequently form amino acids. The experiments on atmospheric synthesis are relevant to us here, whatever the predominant atmospheric conditions may have been, because they illustrate the generality of the Strecker synthesis. One reason why meteorite compositions are similar to that produced in atmospheric discharge experiments is that Strecker synthesis may occur in the parent bodies of the meteorites (Wolman *et al.* 1972).

Hydrothermal vents are local environments on Earth that are often proposed as sites for the origin of life (Baross and Hoffman, 1985; Amend and Shock, 1998). Vents are sources of reducing gases ($H_2$, $H_2S$ *etc.*) that might drive synthesis of organic molecules. They are also habitats for current ecological communities where primary production is chemosynthetic not photosynthetic. Studies of amino acid formation in hydrothermal conditions have been reported. Columns H1 (Marshall, 1994) and H2 (Hennet *et al.* 1992) give the number of times that the amino acid was observed in greater than trace amounts in a large number of experiments. There may be some doubt as to whether the conditions used in these experiments are realistic simulations of what would occur in real hydrothermal vents. Indeed, there is still a debate as to whether amino acid synthesis can occur at all in such conditions (Miller and Bada, 1988; Bada *et al.* 1995). However, these are the best available experimental data that support the argument for hydrothermal synthesis, and we therefore include them in our analysis as we do not want to bias our data choice with respect on one theory or another.

The final three columns in Table 1 are miscellaneous chemical synthesis experiments: S1 is shock synthesis from a gaseous mixture (Bar-Nun *et al.* 1970); S2 is synthesis from ammonium cyanide (Lowe *et al.* 1963, Exp. 2); S3 is synthesis from CO, $H_2$ and $NH_3$ at high temperature in the presence of catalysts (Yoshino *et al.* 1971). This column shows the number of positive identifications of the amino acid from many experiments. These additional experiments are included to illustrate the diversity of conditions that have been proposed as relevant for prebiotic synthesis of amino acids. In choosing the data sets in Table 1, we have tried to include a wide range of points of view, and to emphasize that the nature of the amino acids that are formed in this diverse collection of data is surprisingly consistent.

We have only included data on the twenty biological amino acids that are used in modern proteins. We have not included non-biological amino acids, even though some of these occur in meteorites and spark discharge experiments at relatively high concentrations. Our aim is to understand the order of addition of the biological amino acids to the genetic code, and the non-biological amino acids are not relevant for this question. Also, our argument below relies on comparison with thermodynamic data, which is not available for the non-biological amino acids. In the final section of the paper, we return to the question of why the non-biological amino acids were not incorporated into the code.

**Results - Consensus on Prebiotic Amino Acid Abundances**

We use two independent measures to ascertain amino acid frequencies; a ranking procedure ($R_{obs}$), as well as relative concentrations ($C_{rel}$). In our first approach, amino acids were ranked according to each criterion in Table 1. For example, for criterion M2, glycine (G) is most frequent and is given rank 1. The next two, D and E, are equally frequent, and are both given rank 2.5. The next three are A, V and P, with ranks 4, 5 and 6. All the remaining amino acids are not observed, and are given an equal bottom rank 13.5 (the average of the numbers between 7 and 20). The mean of these ranks from the 12 observations is termed $R_{obs}$. There is good evidence for prebiotic synthesis of all the amino acids down as far as T on the list. The following four (KFRH) are not found in the Miller experiments or in meteorites and occur only in one or two of the other experiments. The remaining amino acids (NQCYMW) are not formed in any of the experiments. The data are not sufficient to make a definite distinction between KFRH and NQCYMW. Our ranking, $R_{obs}$, is close to the ranking from Table V of Trifonov (2004), which we term $R_{code}$ (see Table 2), as it is based on criteria related to genetic code evolution. The top three are the same, and nine amino acids are in the top ten of both orders. We prefer $R_{obs}$ on the grounds that it is derived directly from experimental observables. We will refer to the first ten amino acids in our ranking as the early group, and the bottom ten as the late group. Our hypothesis is that the early amino acids were frequent in the environment at the time of the origin of the genetic code and the first proteins, but that the late amino acids were rare or absent and were only incorporated into proteins when biochemical pathways arose to synthesize them.

Amend and Shock (1998) calculated $\Delta G_{surf}$, the free energy of formation of the amino acids from $CO_2$, $NH_4^+$, and $H_2$ in surface seawater conditions (18°C, 1 atmosphere). Figure 1 reveals the intriguing result that $R_{obs}$ is strongly correlated with $\Delta G_{surf}$ for the 10 early amino acids (r = 0.96). The late amino acids have significantly higher $\Delta G_{surf}$ than the early ones. The means and standard deviations of the early and late groups are 169.3 ± 42.0 and 285.0 ± 94.1 kJ/mol.

Table 2 also lists the molecular weight of each amino acid and the number of ATP molecules required to synthesize the amino acids using the biochemical pathways in *E. coli* (Akashi and Gojobori, 2002). Amino acid biosynthesis pathways branch off from the central pathways of sugar metabolism in the cell (glycolysis and the citric acid cycle). Akashi and Gojobori (2002) calculated the average ATP cost for growth on glucose, acetate and malate as carbon sources, assuming availability of ammonia and sulphate as sources of nitrogen and sulphur. The means and standard deviations of MW for early and late groups are 116.2 ± 20.6 and 157.3 ± 23.2 Da, and the figures for ATP cost are 18.5 ± 6.9 and 36.2 ± 17.3. It is clear that the late group are larger and more thermodynamically costly. We have omitted the two sulphur-containing amino acids, Met and Cys, from Figure 1. Several of the experiments in Table 1 do not include sulphur in the reaction mixture, so these amino acids are bound to be absent from the products. Thus, the value of $R_{obs}$ for Met and Cys is unclear. Nevertheless, Met and Cys are absent from the meteorites, and are late according to $R_{code}$, so it seems reasonable to classify them in the late group. The central thermodynamic point is that the formation reactions are endergonic ($\Delta G_{surf} > 0$). Those with the smallest $\Delta G_{surf}$ should therefore be formed most easily, as they require the least free energy input, as is seen in Figure 1.

Our second measure of amino acid abundances is simply the mean value, $C_{rel}$, of the relative concentrations of the amino acids for 9 data sets (excluding H1, H2 and S3). This is shown in Table 2. Figure 2 shows that $C_{rel}$ drops off roughly exponentially with $\Delta G_{surf}$. This explains why the amino acids with high $\Delta G_{surf}$ are not seen in experiment: their concentration would be too low to detect. The line in Figure 2 is the exponential

$$C_{rel} = 15.8 \exp(-\Delta G_{surf} / 31.3),$$

which is the best exponential fit through the first nine amino acids (excluding T as it is an outlier). An exponential curve would be expected based on thermal equilibrium. Given that the ideal gas constant is 8.3 J/mol-K, 31.3 kJ corresponds to a temperature of $3.7 \times 10^3$ K. However, the exponential dependence may also be a result of the kinetics of the formation process, with progressively more complex molecules having slower formation rates, in which case the exponential decay constant cannot be interpreted as a temperature. Isotopic studies of the Murchison meteorite suggest the importance of the formation kinetics (Pizzarello *et al.* 2004). Another important caveat is that amino acids may differ in their long-term stability, so that relative concentrations seen in the meteorites will depend on the rates of decomposition to some degree.

The free energy of formation under hydrothermal conditions, $\Delta G_{hydro}$ (100$^o$C, 250 atmospheres) was also calculated by Amend and Shock (1998). Formation of some of the amino acids is exergonic ($\Delta G_{hydro} < 0$) under hydrothermal conditions, and even the endergonic ones are less positive than they are at the surface. This was used to support the idea that the first organisms might have been deep-sea chemosynthesizers. However, we find that there is no correlation between $R_{obs}$ and $\Delta G_{hydro}$, as seen in Figure 3. The highest ranking amino acids are intermediate in $\Delta G_{hydro}$, the late amino acids span the full range from lowest to highest $\Delta G_{hydro}$, and there is no significant difference in $\Delta G_{hydro}$ between early and late amino acids.

Although we would not expect $\Delta G_{hydro}$ to be a predictor of what amino acids are seen in meteorites or the atmospheric discharge experiments, it is puzzling that it also does not predict which amino acids are seen in the two experiments designed to simulate hydrothermal systems (H1 and H2). In fact, these experiments give results that agree fairly well with the combined ranking from the other data, rather than the calculated $\Delta G_{hydro}$. This may be an indication that relative concentrations in these experiments are controlled by kinetic factors and that the simplest amino acids are quickest to form, even though some of the larger amino acids have much lower $\Delta G_{hydro}$.

Another surprise is that thermodynamic calculations in surface seawater conditions make useful predictions about the amino acids in meteorites. This may be an indication that water-based chemistry is important also in the meteorites. Indeed, it has been argued (Martins *et al.* 2007) that the organic content of meteorites is strongly influenced by aqueous alteration inside the meteorite parent body, and that differences in this alteration may account for the large differences in the content found among carbonaceous chondrites.

It is known that modern organisms are sensitive to the thermodynamic costs of amino acid synthesis. It has been observed that amino acids with lower ATP costs of synthesis have higher frequencies in highly expressed proteins (Akashi and Gojobori, 2002; Seligman, 2003). There is also some suggestion that the differences between synthesis costs in hyperthermophiles and mesophiles may cause different amino acids to be preferred in high expression proteins in hyperthermophiles (Swire, 2007). There are definitely differences in amino acid frequencies between proteins in hyperthermophiles and their mesophile relatives, but this can arise from selection for stability of proteins at high temperatures (Berezovsky *et al.* 2005) and does not necessarily indicate selection to economize on amino acid synthesis.

**Discussion - Implications for the origin of the genetic code and protein evolution**

The range of different specific functions to which proteins have adapted in modern organisms is astounding. As far as we know, proteins cannot achieve replication and heredity by themselves. In modern organisms, proteins are synthesized by translation, which is crucially dependent on mRNAs, rRNAs and tRNAs. This is the main reason for believing that an RNA world existed early in the history of life (Joyce, 2002) in which both genetic and catalytic roles were performed by RNA molecules. It is possible that amino acids or short peptides played some part in the RNA world, or that there were molecules with short peptide sequences attached to nucleotides, like some current coenzymes. However, it was the invention of translation and the genetic code that coupled the hereditary mechanism provided by template-directed replication of nucleic acids to the catalytic possibilities of proteins. Prior to the genetic code, proteins could not evolve.

The standard genetic code is shown in Figure 4. Amino acids are coded by between 1 and 6 codons, with most amino acids being coded by either 2 or 4 codons. It is known that this standard code evolved prior to the common ancestor of all current organisms, and it has remained unchanged since then in the majority of genomes of archaea, bacteria and eukaryotes. There are numerous cases of codon reassignments that have occurred much more recently in specific lineages of mitochondrial genomes and some others (Knight *et al.* 2001; Sengupta *et al.* 2007). However, it is the early phase of code evolution leading to the establishment of the standard code that is relevant to the current paper. Early versions of the code probably used a smaller repertoire of amino acids, each coded by a larger number of codons. The codon table was divided up into progressively smaller blocks as successive amino acids were added. This was suggested as far back as Crick (1968). This idea is also central to the coevolution theory for the genetic code (Wong, 1975; 2005; Di Giulio and Medugno, 1999; Di Giulio, 2005) which considers addition of amino acids to the code in an order that is determined by the way the amino acid biosynthetic pathways evolved. The coevolution theory proposes the same set of early amino acids as we do, based on the argument that these are synthesized in Miller's experiments, and because the simplest amino acids are found at the beginnings of the biosynthetic pathways. The strong correlation between $R_{obs}$ and $\Delta G_{surf}$ found above and also the close agreement with $R_{code}$ (Trifonov, 2004) suggest that our ranking is a meaningful prediction of the relative frequencies of abiotically synthesized amino acids that would have been available to the first organisms. Our results support the coevolution theory in that they show that none of the proposed mechanisms of non-biological synthesis is able to synthesize the late amino acids; hence the late amino acids could only have been incorporated into the code after biochemical synthesis pathways arose. Conversely, however, our results show that the early amino acids can be synthesized by several non-biological means, and were likely to have been available in the environment. Hence, biosynthetic pathways were not necessarily relevant at the very early stages on code evolution when only a small number of amino acids were encoded.

Another key point of the coevolution theory is that the precursor-product relationships between amino acids determine the positions in the code into which new amino acids are assigned. Specifically, new amino acids are assigned to codons that previously belonged to their precursors (Wong, 1975; Di Giulio and Medugno, 1999). In our own view, however, it is the physical properties of the amino acids that have a greater influence on which amino acids are assigned to which codons. The layout of the code is far from random. It is important to note that codons that differ by only one out of the three base positions are often assigned to amino acids with similar properties. As a result, the effects of mutations and translational errors are minimized. The canonical genetic code is better than all but a tiny fraction of randomly rearranged codes in this respect (Freeland *et al.* 2003). This suggests the strong role of natural selection in building up the code. Selection will favour the assignment of new amino acids to codons that previously coded for amino acids with similar properties (Higgs and Pudritz, 2007; Higgs 2009) because this will be minimally disruptive to the genes that were already encoded by the smaller amino acid set before the new one was introduced. As a result, neighbouring codons in the final code will end up with similar properties.

It is well known that hydrophobicity of the amino acids correlates with the column in the code rather than the row (Jungck, 1978; Taylor and Coates, 1989). A principle component analysis of amino acid physical properties (Urbina *et al.* 2006) shows that amino acids in the same column cluster in property space, and that this has observable consequences for the rate of sequence evolution at first and second codon positions and for the degree to which amino acid frequencies in proteins are influenced by biased mutation at the DNA level. We suggest that the current columnar structure in the current standard code has been retained from the very earliest stages of code evolution. An intriguing observation in this regard is that the five earliest amino acids according to our ranking (Gly, Ala, Asp, Glu, Val) appear on the bottom row of the genetic code table (first position base G). This suggests to us that the second base was the most important discriminator in the early code. The early code may have had a four-column structure in which all codons with the same second base would have coded for the same amino acid: NUN for Val, NCN for Ala, NAN for Asp or Glu and NGN for Gly (where N denotes any base). Selection would have favoured addition of further amino acids into particular columns of the code table according to physicochemical properties, so that hydrophobic amino acids end up in the Val column, hydrophilic amino acids end up in the Asp/Glu column *etc*. Higgs (2009) has considered the way selection would favour the pathways of code evolution from the four-column code to the current standard code.

Another related idea is the two-out-of-three argument of Lagerkvist (1978; 1986), which was recently extended by Lehmann and Libchaber (2008). This proposes that the wobble base pair (first anticodon base with the third codon base) is much less important that the other two pairs in the codon-anticodon interaction. The usual wobble rules allow a tRNA with a U at the wobble position to translate codons ending A and G, and a tRNA with G at the wobble position to translate codons ending U and C. Hence, two tRNAs are needed to translate a four-codon family. According to the two-out-of-three argument, there is also a significant probability that the wobble-U tRNA can translate all four codons, provided that the pairing at the first and second positions is strong. Thus, when there is strong pairing at the first two positions, the four-codon box must be assigned to a single amino acid, whereas when the pairing is weak, the two tRNAs do not translate the same codons, so it is possible to split a four-codon box into two groups of two codons. This argument potentially gives an explanation of the positions of the two-codon and four-codon families in the code.

This leads to the question of how many amino acids can be encoded by a system with 64 codons and why there are 20 amino acids in the standard code. There should be a selective advantage of adding another amino acid to the code that comes from the increased diversity of proteins that can be made with an increased amino acid repertoire. Nevertheless, several amino acids are found to be as frequent in meteorites as the early biological ones and yet did not become incorporated into the code (Lu and Freeland, 2006; Wiltschi and Budisa, 2007). Many of the four-codon boxes are split between two amino acids. If this were done in every four-codon box, there could be up to 32 amino acids specified. On the other hand, if we accept the two-out-of-three argument, then there are 8 boxes which cannot be split. This reduces the maximum number of amino acids to 24 (8 four-codon families and 16 two-codon families). Splitting a two codon family into separate single codons is possible in the case of Met and Trp codons, but this does not increase the number of amino acids encoded in either case, and in many mitochondrial codes, Met and Trp both have two codons (Sengupta *et al.* 2007). None of these arguments explains why there should be some amino acids with 6 codons. Thus, the fact that the observed 20 amino acids is fewer than the maximum seems to require some explanation.

It may be that if an amino acid is too chemically similar to others that are already in the code, there is little selective advantage to adding it. Furthermore, adding more amino acids requires reducing the size of the codon blocks. Hence the probability of translational error becomes larger. As the number of amino acids in the code increases, the advantage of adding a further amino acid becomes small relative to the disadvantage of increased translational error rate (Higgs, 2009). This would explain why the process of code evolution stopped with an intermediate number of amino acids incorporated. For some of the non-biological amino acids that are seen in meteorites and prebiotic synthesis experiments, chemical reasons have been proposed as to why they were avoided. For example, Weber and Miller (1981) argue that amino acids without an α hydrogen (such as α-amino isobutyric acid and isovaline) are avoided because there would be increased steric hindrance and lack of flexibility of the backbone. However, they were unable to propose a reason for avoidance of α-amino-n-butyric acid, norvaline and norleucine. These are similar in properties to valine, leucine and isoleucine, so it may simply be that the existing set of 20 gives a fairly complete coverage of physical property space already.

In addition to the role of physical properties and of precursor product relationships, another factor that has also been proposed to influence the initial codon assignments is that there may have been direct interactions between RNA triplets and amino acids (Yarus *et al.* 2005). Despite their differences, all these arguments agree that early proteins were composed of a small set of amino acids. Selection would favour the addition of new amino acids because the number of possible protein sequences increases tremendously with each amino acid in the repertoire. However, proteins composed of a restricted set of amino acids must be functional at each step of this process in order for selection to favour the use of proteins. Several studies have shown that sequences composed of small sets of amino acids can have structures and functions similar to those of 20-amino acid proteins (Davidson *et al.* 1995; Riddle *et al.* 1997; Babajide *et al.* 1997; Doi *et al.* 2005). Thus selection may favour the increasing use of more diverse and better adapted proteins and hence the RNA world would be eventually overtaken by the protein-dominated world.

In summary, we have shown that in spite of the conflicting opinions on the mechanisms and locations of molecular synthesis, thermodynamics cuts across these distinctions and predicts which amino acids are formed most easily. Our results also indicate that a certain degree of universality would be expected in the types of organic molecules seen on other earth-like planets. Should life exist elsewhere, it would not be surprising if it used at least some of the same amino acids we do. Simple sugars, lipids and nucleobases might also be shared. Our analysis suggests that the first genetic code was a stripped down version of our present code, one whose simpler structure reflected the limited set of available amino acids in the prebiotic environment. Although there are countless ways that the code could have developed from those origins, the combined actions of thermodynamics and subsequent natural selection suggest that the genetic code we observe on the Earth today may have significant features in common with life throughout the cosmos.

**Acknowledgements**

This work is supported by Canada Research Chairs and the Natural Sciences and Engineering Research Council of Canada. We thank Chris McKay for useful suggestions on this work, and two anonymous referees, whose comments were extremely helpful.

Table 1

Relative concentrations of amino acids observed in non-biological contexts. With the exception of columns H1, H2 and S3, the quoted figures are concentrations normalized such that Gly = 1.0. For columns H1, H2 and S3, the figures are the number of experiments in which the amino acid was observed in greater than trace amounts. M denotes meteorites, I denotes icy grains, A denotes atmospheric synthesis, H denotes hydrothermal synthesis, and S denotes other chemical syntheses (details of sources are given in the text). $R_{obs}$ is the mean rank derived from these observations. The final 6 amino acids are not observed

|  | M1 | M2 | M3 | I1 | A1 | A2 | A3 | H1 | H2 | S1 | S2 | S3 | $R_{obs}$ |
|---|---|---|---|---|---|---|---|---|---|---|---|---|---|
| G - Gly | 1.00 | 1.0 | 1.000 | 1.000 | 1.000 | 1.000 | 1.000 | 18 | 12 | 1.000 | 1.000 | 40 | 1.1 |
| A - Ala | 0.34 | 0.4 | 0.380 | 0.293 | 0.540 | 1.795 | 0.155 | 15 | 8 | 0.473 | 0.097 | 20 | 2.8 |
| D - Asp | 0.19 | 0.5 | 0.035 | 0.022 | 0.006 | 0.077 | 0.059 | 10 | 10 | 0.000 | 0.581 | 30 | 4.3 |
| E - Glu | 0.40 | 0.5 | 0.110 | 0.000 | 0.010 | 0.018 | 0.000 | 6 | 11 | 0.000 | 0.000 | 20 | 6.8 |
| V - Val | 0.19 | 0.3 | 0.100 | 0.012 | 0.000 | 0.044 | 0.000 | 1 | 0 | 0.006 | 0.000 | 2 | 8.5 |
| S - Ser | 0.00 | 0.0 | 0.003 | 0.072 | 0.000 | 0.011 | 0.018 | 8 | 11 | 0.000 | 0.154 | 0 | 8.6 |
| I - Ile | 0.13 | 0.0 | 0.060 | 0.000 | 0.000 | 0.011 | 0.000 | 8 | 9 | 0.000 | 0.002 | 4 | 9.1 |
| L - Leu | 0.04 | 0.0 | 0.035 | 0.000 | 0.000 | 0.026 | 0.000 | 3 | 0 | 0.001 | 0.002 | 7 | 9.4 |
| P - Pro | 0.29 | 0.1 | 0.000 | 0.001 | 0.000 | 0.003 | 0.000 | 9 | 0 | 0.000 | 0.000 | 2 | 10.0 |
| T - Thr | 0.00 | 0.0 | 0.003 | 0.000 | 0.000 | 0.002 | 0.000 | 2 | 0 | 0.000 | 0.002 | 1 | 11.7 |
| K - Lys | 0.00 | 0.0 | 0.000 | 0.000 | 0.000 | 0.000 | 0.000 | 7 | 0 | 0.000 | 0.000 | 14 | 12.6 |
| F - Phe | 0.00 | 0.0 | 0.000 | 0.000 | 0.000 | 0.000 | 0.000 | 4 | 0 | 0.000 | 0.000 | 1 | 13.2 |
| R - Arg | 0.00 | 0.0 | 0.000 | 0.000 | 0.000 | 0.000 | 0.000 | 0 | 0 | 0.000 | 0.000 | 15 | 13.3 |
| H - His | 0.00 | 0.0 | 0.000 | 0.000 | 0.000 | 0.000 | 0.000 | 0 | 0 | 0.000 | 0.000 | 15 | 13.3 |
| NQCYMW | 0.00 | 0.0 | 0.000 | 0.000 | 0.000 | 0.000 | 0.000 | 0 | 0 | 0.000 | 0.000 | 0 | 14.2 |

Table 2
Thermodynamic and evolutionary properties of amino acids. $R_{obs}$ = ranking derived from Table 1; $R_{code}$ = ranking from Trifonov (2004); $C_{rel}$ = mean relative concentration from Table 1; $\Delta G_{surf}$ and $\Delta G_{hydro}$ = free energies of formation in kJ/mol (Amend and Shock, 1998); ATP cost of synthesis (Akashi and Gojobori, 2002); MW = molecular weight in Da.

|         | $R_{obs}$ | $R_{code}$ | $C_{rel}$ | $\Delta G_{surf}$ | $\Delta G_{hydro}$ | ATP  | MW    |
|---------|-----------|------------|-----------|--------------------|---------------------|------|-------|
| G - Gly | 1.1       | 3.5        | 1.0000    | 80.49              | 14.89               | 11.7 | 75.1  |
| A - Ala | 2.8       | 4.0        | 0.4970    | 113.66             | -12.12              | 11.7 | 89.1  |
| D - Asp | 4.3       | 6.0        | 0.1633    | 146.74             | 32.78               | 12.7 | 133.1 |
| E - Glu | 6.8       | 8.1        | 0.1153    | 172.13             | -1.43               | 15.3 | 147.1 |
| V - Val | 8.5       | 6.3        | 0.0724    | 178.00             | -70.12              | 23.3 | 117.1 |
| S - Ser | 8.6       | 7.6        | 0.0286    | 173.73             | 69.47               | 11.7 | 105.1 |
| I - Ile | 9.1       | 11.4       | 0.0226    | 213.93             | -96.4               | 32.3 | 131.2 |
| L - Leu | 9.4       | 9.9        | 0.0116    | 205.03             | -105.53             | 27.3 | 131.2 |
| P - Pro | 10.0      | 7.3        | 0.0437    | 192.83             | -38.75              | 20.3 | 115.1 |
| T - Thr | 11.7      | 9.4        | 0.0008    | 216.50             | 53.51               | 18.7 | 119.1 |
| K - Lys | 12.6      | 13.3       | 0         | 258.56             | -28.33              | 30.3 | 146.2 |
| F - Phe | 13.2      | 14.4       | 0         | 303.64             | -114.54             | 52.0 | 165.2 |
| R - Arg | 13.3      | 11.0       | 0         | 409.46             | 197.52              | 27.3 | 174.2 |
| H - His | 13.3      | 13.0       | 0         | 350.52             | 154.48              | 38.3 | 155.2 |
| N - Asn | 14.2      | 11.3       | 0         | 201.56             | 83.53               | 14.7 | 132.1 |
| Q - Gln | 14.2      | 11.4       | 0         | 223.36             | 44.04               | 16.3 | 146.1 |
| C - Cys | 14.2      | 13.8       | 0         | 224.67             | 60.24               | 24.7 | 121.2 |
| Y - Tyr | 14.2      | 15.2       | 0         | 334.20             | -59.53              | 50.0 | 181.2 |
| M - Met | 14.2      | 15.4       | 0         | 113.22             | -174.71             | 34.3 | 149.2 |
| W - Trp | 14.2      | 16.5       | 0         | 431.17             | -38.99              | 74.3 | 204.2 |

Figure Captions.

Figure 1 – The ranking $R_{obs}$ for the early amino acids (circles) correlates well with the free energy of formation in surface seawater conditions, $\Delta G_{surf}$. The late amino acids (triangles) have high $\Delta G_{surf}$.

Figure 2 – The mean relative concentration $C_{rel}$ for the early amino acids decreases approximately exponentially with $\Delta G_{surf}$.

Figure 3 – There is no correlation between the ranking $R_{obs}$ and the free energy of formation in hydrothermal conditions, $\Delta G_{hydro}$, for either the early or late amino acids.

Figure 4 – The table of assignments between codons and amino acids in the standard genetic code.

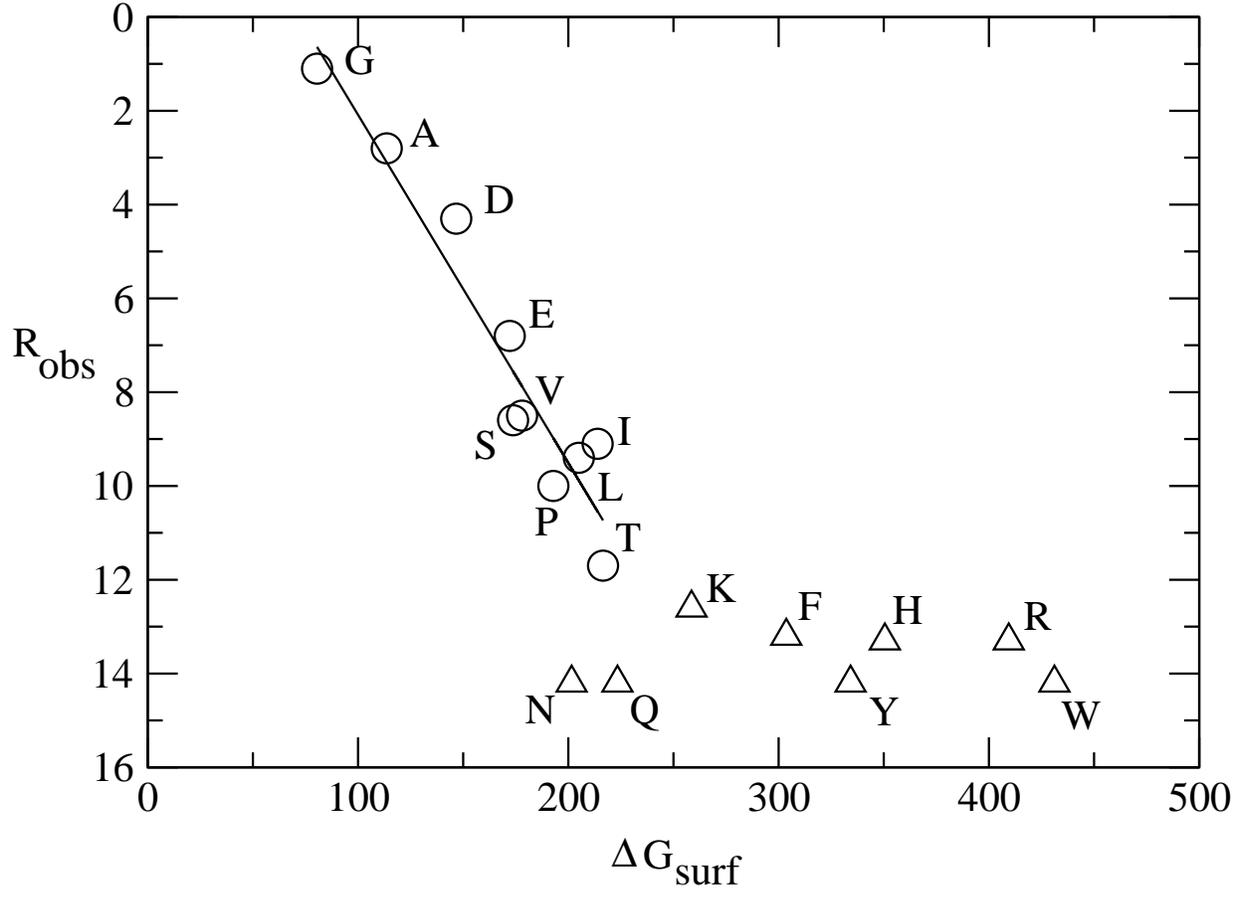

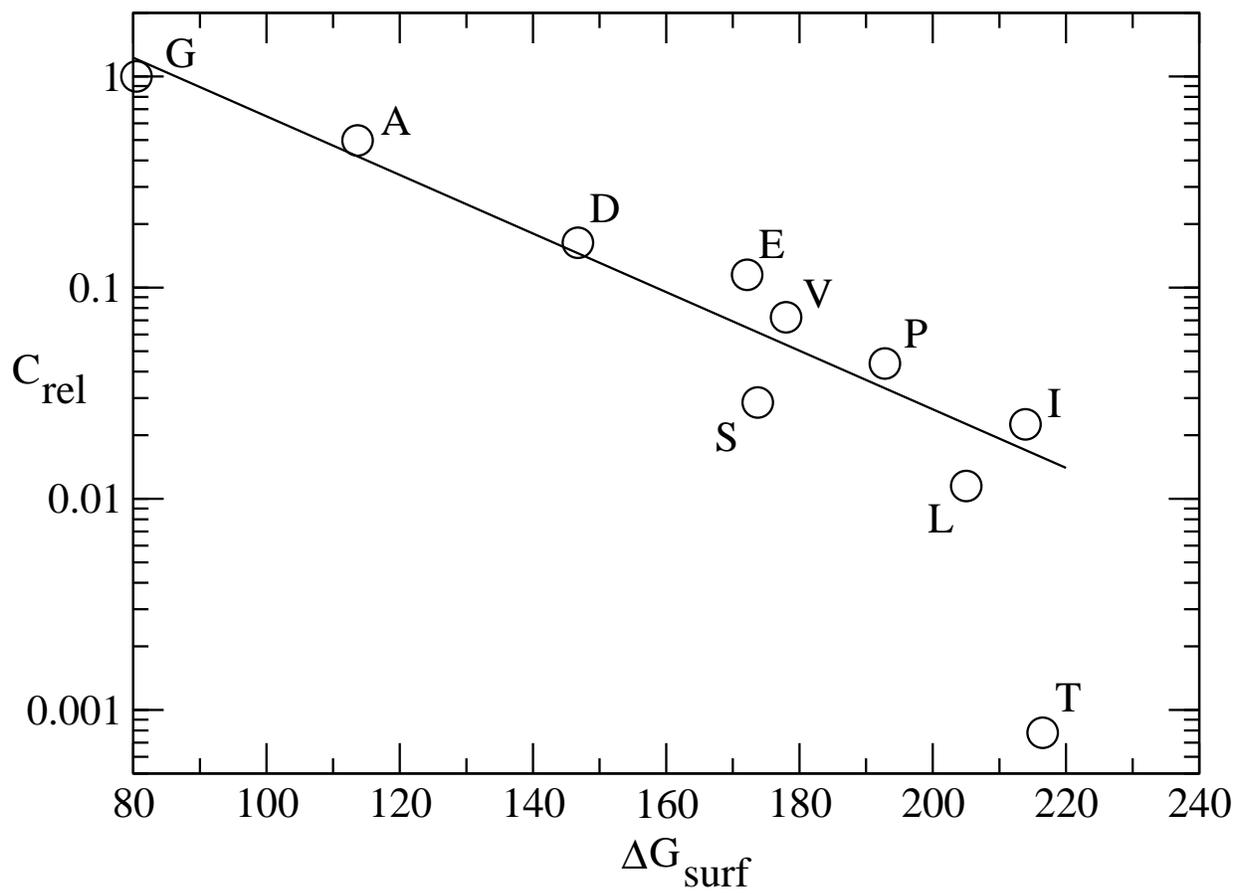

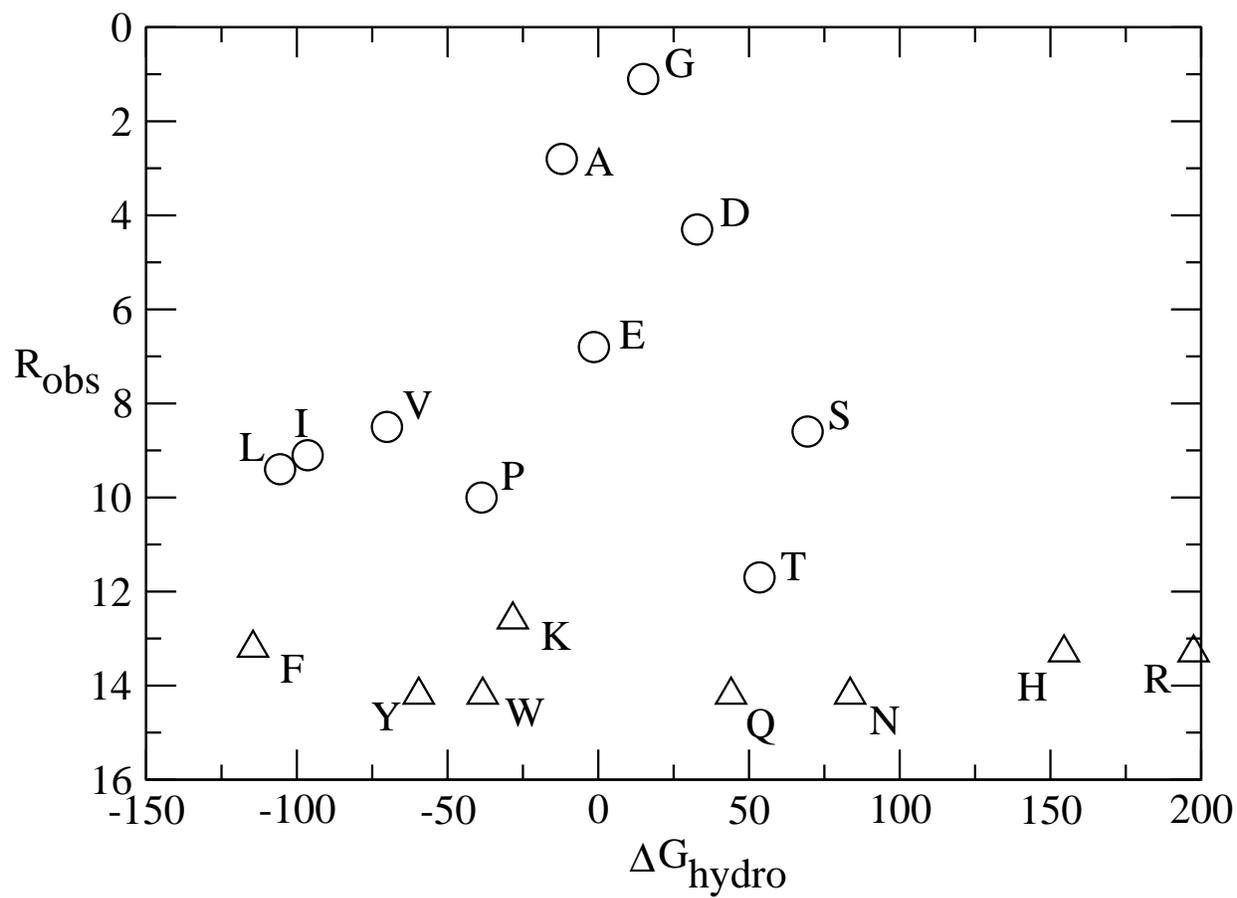

Fig 4

|   | U | C | A | G |   |
|---|---|---|---|---|---|
| U | UUU Phe<br>UUC Phe<br>UUA Leu<br>UUG Leu | UCU Ser<br>UCC Ser<br>UCA Ser<br>UCG Ser | UAU Tyr<br>UAC Tyr<br>UAA Stop<br>UAG Stop | UGU Cys<br>UGC Cys<br>UGA Stop<br>UGG Trp | U<br>C<br>A<br>G |
| C | CUU Leu<br>CUC Leu<br>CUA Leu<br>CUG Leu | CCU Pro<br>CCC Pro<br>CCA Pro<br>CCG Pro | CAU His<br>CAC His<br>CAA Gln<br>CAG Gln | CGU Arg<br>CGC Arg<br>CGA Arg<br>CGG Arg | U<br>C<br>A<br>G |
| A | AUU Ile<br>AUC Ile<br>AUA Ile<br>AUG Met | ACU Thr<br>ACC Thr<br>ACA Thr<br>ACG Thr | AAU Asn<br>AAC Asn<br>AAA Lys<br>AAG Lys | AGU Ser<br>AGC Ser<br>AGA Arg<br>AGC Arg | U<br>C<br>A<br>G |
| G | GUU Val<br>GUC Val<br>GUA Val<br>GUG Val | GCU Ala<br>GCC Ala<br>GCA Ala<br>GCG Ala | GAU Asp<br>GAC Asp<br>GAA Glu<br>GAG Glu | GGU Gly<br>GGC Gly<br>GGA Gly<br>GGG Gly | U<br>C<br>A<br>G |